# Time-resolved Fourier transform intracavity spectroscopy with a Cr$^{2+}$:ZnSe laser

**Nathalie Picqué, Fatou Gueye and Guy Guelachvili**
*Laboratoire de Photophysique Moléculaire, Unité Propre du C.N.R.S., Bâtiment 350, Université de Paris-Sud, 91405 Orsay, France*

**Evgeni Sorokin and Irina T. Sorokina**
*Institut für Photonik, TU Wien, Gusshausstr. 27/387, A-1040 Vienna, Austria*

Corresponding author:
Dr. Nathalie Picqué,
Laboratoire de Photophysique Moléculaire
Unité Propre du CNRS, Université de Paris-Sud, Bâtiment 350
91405 Orsay Cedex, France
Website: http://www.laser-fts.org
Phone nb: 33 1 69 15 66 49
Fax nb: 33 1 69 15 75 30
Email: nathalie.picque@ppm.u-psud.fr

**Abstract:**
Intracavity laser absorption spectroscopy (ICLAS) with an evacuated $Cr^{2+}$: ZnSe laser is performed with a high-resolution time-resolved Fourier transform interferometer with a minimum detectable absorption coefficient equal to $4\ 10^{-9}$ cm$^{-1}$ Hz$^{-1/2}$ in the 2.5µm region. This represents the extreme limit presently reached in the infrared by ICLAS with Doppler limited resolution. The broad gain band of the crystal allows a spectral coverage at most equal to 125 nm, wide enough to see entire vibration bands. Weak $CO_2$ bands observed up to now only in the Venus atmosphere are recorded for the first time in a laboratory. $H_2O$ detection limit down to 0.9 ppbv is also demonstrated.

High sensitivity spectroscopic detection is of special importance for trace gas sensing applications which have an increasing impact in numerous areas including fundamental spectroscopy, atmospheric chemistry, industrial process, and medical diagnostics. The infrared spectral region has the considerable advantage to be the location of relatively strong absorption molecular bands. These allow a better minimum detectable absorption coefficient, approximated as the inverse of the product $L$ x $SNR$, where $L$ is the absorption length and $SNR$ is the signal to noise ratio of the measurement. Detection limits down to sub ppb (parts-per-billion) have been achieved with photoacoustic[1,2] and cavity ringdown[3] sensitive techniques. The tunable lasers used by these techniques do not allow simultaneous coverage of broad spectral ranges. More importantly, broadband detection is required when there is a need to selectively detect several molecular gas species simultaneously by means of high sensitivity absorption spectroscopy.

Intracavity laser absorption spectroscopy (ICLAS)[4] has been known for a long time as a useful approach to the absorption technique. It provides one of the highest sensitivity with the additional advantage of a simultaneous coverage of a broad spectral range similar to the laser gain bandwidth. ICLAS consists of placing the absorbing sample inside a laser cavity, the gain of which is broader than the sample absorption lines. Since only broadband losses are compensated by the laser gain, the laser operates for the absorption lines like a multipass cell. The equivalent absorption path length $L$ is equal to the product of the velocity of light $c$ by the generation time $t_g$, which separates the beginning of the laser pulse from the observation. Various laser gain media have been used mainly for applications in the visible range.

The infrared region is still poorly explored with the ICLAS method, due to the lack of efficient lasers and spectroscopic instruments. In the infrared region beyond 2.5 µm only one ICLAS experiment has been demonstrated to our knowledge.[5] Using a KCl:Li Fa(II) color center laser, residual $H_2O$ and $CO_2$ atmospheric lines were obtained in the 2.636-2.640 µm region, with a minimum detectable absorption coefficient equal to 3 $10^{-7}$ $cm^{-1}$.

For the development of ICLAS in the infrared, the $Cr^{2+}$:ZnSe and the other $Cr^{2+}$-chalcogenide laser materials appear promising. They offer wide gain bands, centered in the mid-IR regions between 2 and 3.5 µm.[6] Additionally, they operate at room temperature and can be pumped with available solid-state, fiber, and diode lasers. Recently, first use of $Cr^{2+}$:ZnSe laser for ICLAS was performed using a pulsed Co:$MgF_2$ pump laser.[7] The recorded spectra of the ambient atmosphere were oversaturated by the ambient water vapor. Furthermore, the 2.4 µm IR light had to be up-converted to enable detection by a grating spectrometer operating in the visible range, where linear CCD arrays are conveniently available.

For analyzing the broadband laser emission spectrum, Fourier transform (FT) spectrometers appear as the most efficient instruments. They need only one single detector. Furthermore, time-resolved FT (TRFT) interferometers are available, which enable the temporal sampling of the laser transient spectral dynamics observed in the repetitive pulsed operation mode. This results in the measurement of numerous spectra observed at different generation times under the same experimental sample conditions. This allows precise determination of equivalent absorbing paths by verifying the consistency of time evolution of different spectral lines, enabling the accurate line intensity measurement by ICLAS.

In this Letter, we report the first spectrum recorded at the extreme infrared limit ever reached by ICLAS with Doppler-limited resolution. It results from the combination of a broadband $Cr^{2+}$ZnSe laser and a time-resolved FT interferometer[8]. Thanks to the sensitivity provided by the intracavity path length enhancement, we are able to observe for the first time in laboratory conditions weak difference bands of carbon dioxide, which were up to now only seen in

planetary spectra.[9]

The experimental intracavity set up is schematically shown on Fig. 1. An $Er^{3+}$-fiber laser pumped $Cr^{2+}$: ZnSe laser is used as a strongly multimode ICLAS source. An acousto-optic modulator (AOM) with 0.5 µs rise-time chops the pumping beam. The 8 mm long single-crystal diffusion-doped $Cr^{2+}$: ZnSe Brewster cut sample[10] absorbs about 85% of pump radiation at 1607 nm. The crystal has a peak emission cross-section equal to 1.1 x $10^{-18}$ $cm^2$ at 2450 nm. The gain spectrum full-width at half-maximum is as broad as 860 nm. The X-fold four-mirror cavity configuration is optimized to avoid the possible parasitic étalon effects. The standing wave cavity consists of two highly reflective (HR) concave mirrors, a HR plane mirror and a plane output coupler (OC). In the 2300-2700 nm region, the HR reflection coefficients are 99.8% whereas the OC has 6% transmission. In this configuration, the threshold pump power is 160 mW. In the present experiments, we kept the pump parameter η, ie the ratio of the pump power to its threshold value, at η=1.6, and the pump pulse was 400 µs long at 2 kHz repetition rate. From the monitoring of the total laser intensity on an InSb detector, the pulse-to-pulse variability of the optimized laser was found to be negligible and was consequently not taken care of. The light leaking outside the laser cavity through the OC is sent into the interferometer. The $Cr^{2+}$: ZnSe laser is able to operate under secondary vacuum. The vacuum chamber windows (not shown on the diagram) are made of BK7 on the optical pumping laser side and of $CaF_2$ on the interferometer side.

The stepping-mode time-resolved interferometer is equipped with $CaF_2$ beam-splitter and two InSb detectors cooled with liquid nitrogen. In time-resolved approach, the operating mode is such that, at each fixed path difference value, a set of *n* time samples is recorded, while the path-difference is held constant. This time sample acquisition is repeated identically at each path-difference step, up to the maximum value of the path-difference. All time samples are rearranged afterwards to constitute a series of *n* independent interferograms each of them reporting the spectral state of the source at one of the given instant of the time sampling procedure. Finally the TRFT spectrum obtained in one experiment is made of *n* time-component spectra, each of them exhibiting all the usual advantages of Fourier transform spectrometry. When applied to ICLAS, TRFT spectroscopy brings the following advantages. The intracavity gas sample is observed for all time-components, under identical pressure and temperature conditions, which are easily measured due to the small size of the cell. The only varying quantity is then the absorption length *L*. Since *L* often departs [4] from the ideal law $L = c \times t_g$, errors often spoil the ICLAS line intensity measurements. This difficulty is overcome with TRFT spectroscopy, which allows checking the consistency of *L* evolution and making *a posteriori* corrections with multispectrum fitting procedures.

Two series of spectra were recorded. The first ones were obtained with the laser chamber evacuated down to a pressure of 7 $10^{-3}$ Pa. The laser emission is centered at 2490 nm. For intracavity spectroscopy purposes, secondary-vacuum evacuation of the laser chamber appears necessary. Under poorly evacuated chamber conditions, the parasitic absorption of the major air constituents would indeed totally obscure the range of interest in the recorded spectra, as observed in Ref. [7]. Another advantage of the vacuum chamber is the possibility to insert the analyzed molecular sample in the laser cavity, with no need of an optical cell, the windows of which would narrow the laser emission bandwidth. Also the filling ratio of the cavity is kept optimum (close to unity), and additional experimental complexity is avoided. The following recorded molecular spectra were benefiting from these advantages.

The second series was made of several $CO_2$ absorption TRFT spectra recorded under various pressures, from 396 to 7980 Pa and various resolutions, the highest being equal to 6.9 pm (11 $10^{-3}$ cm$^{-1}$) unapodized. An illustration of the TRFT spectrum is plotted on Figure 2 with the resolution reduced to 62 pm (0.1 cm$^{-1}$) for clarity. It is obtained with the chamber filled with 6650 Pa of $CO_2$. $H_2O$ lines are also present as impurities in the spectrum. At each given path difference step, 64 time samples are taken from a given laser pulse with a 0.32 µs time-resolution and 256 pulses are averaged to improve the SNR. The data acquisition procedure results in a 64 time-components spectrum. Each time-component made of 65536 spectral elements covers 452 nm (737 cm$^{-1}$). At small $t_g$, the recorded laser spectral emission covers 125 nm (202 cm$^{-1}$), filling only about one quarter of the available spectral range actually measured by the interferometer. This emission is centered on the much broader laser gain curve. With improved experimental conditions aiming at reducing the broadband cavity losses, one can expect to slow down the laser dynamics and to increase by a factor of 2 the already broad explored spectral range. The spectral narrowing at longer $t_g$ reduces this spectral range by a factor of about 2.5 for the longest $t_g$ equal to about 27 µs (8.1 km equivalent absorbing path).

The explored spectral domain is the location of the two weak vibration-rotation $CO_2$ bands $2\nu_3 - \nu_2$ and $2\nu_3 + \nu_2 - 2\nu_2$. A restricted portion corresponding to the R-branch of the $2\nu_3 - \nu_2$ of two time components at $t_g$ respectively equal to 8.24 and 15.28 µs is shown with an expanded wavelength on Figure 3. The R-branch head at R(59) is seen on both traces. The two strongest lines on the upper trace belong to the $\nu_3$ band of residual $H_2O$. Rovibrational transitions of the $\nu_1$ band of $H_2O$ are also present. The different slopes of the background reveal the effect of spectral narrowing. The highest equivalent absorbing path reached in the $CO_2$ experiment is equal to 8.1 km. With a *SNR* ratio of about 40 the corresponding minimum detectable absorption coefficient is 3.1 $10^{-8}$ cm$^{-1}$. Data recording time is 5 hours. Thanks to Fourier transform spectroscopy, a total of about 4,200,000 spectral elements are measured, among which 1,050,000 cover the laser emission range. Consequently the minimum detectable absorption coefficient at 1-second time averaging is 4 $10^{-9}$ cm$^{-1}$ Hz$^{-1/2}$. The most intense $H_2O$ observed line is located at 2536.3637 nm (3942.652 cm$^{-1}$). Its intensity is[11] 1.662 $10^{-20}$ cm.molecule$^{-1}$. The 3 x $10^{-8}$ cm$^{-1}$ minimum detectable absorption coefficient then corresponds to 0.9 ppbv water vapor detection limit.

In conclusion, we demonstrate an ICLAS experiment covering more than 100 nm spectral range in the 2.5-µm region with Doppler-limited resolution and 3 x $10^{-8}$ cm$^{-1}$ minimum detected absorption, using the time-resolved Fourier spectroscopy technique. Further experiments aiming at increasing both the sensitivity and the simultaneously explored spectral range are being actively pursued.

The authors are grateful to A. A. Kachanov for initiating the contact between the two groups and R. Vasquez for helpful technical assistance. We acknowledge the support from the Wirtschatfskammer Wien and the French-Austrian exchange program Amadeus. N. Picqué's and G. Guelachvili's email addresses are nathalie.picque@ppm.u-psud.fr and guy.guelachvili@ppm.u-psud.fr

Figure Captions

Fig. 1. Schematic diagram of the TRFT-ICLAS experiment. The dashed rectangle represents the vacuum chamber. The signal of the total intensity variation of the laser beam after the output coupler (OC) is also shown on the figure. AOM: acousto-optic modulator.

Fig. 2. $CO_2$ time-resolved spectrum made of 64 time-components. Two consecutive components are 0.32 µs from each other. This corresponds to a 96-meter increase of the equivalent absorbing path $L$. The upper right-hand enclosure gives versus time the total laser intensity. The cavity build up time is 8.1 µs and the relaxation oscillation period is 3.6 µs.

Fig. 3. Restricted portion of two components of the time-resolved spectrum shown on Fig. 2. Line profiles are Doppler limited. Equivalent absorbing path values $L$ are respectively 2.5 and 4.6 km. The present spectra had not previously been recorded under laboratory conditions and could only be observed in the atmosphere of Venus [9] made of 96.5% $CO_2$.

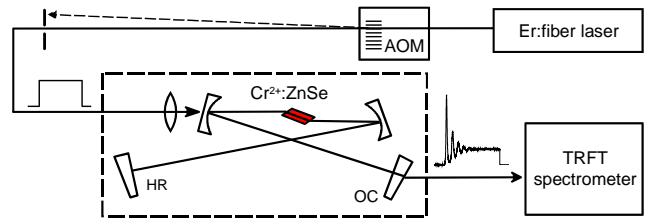

Figure 1

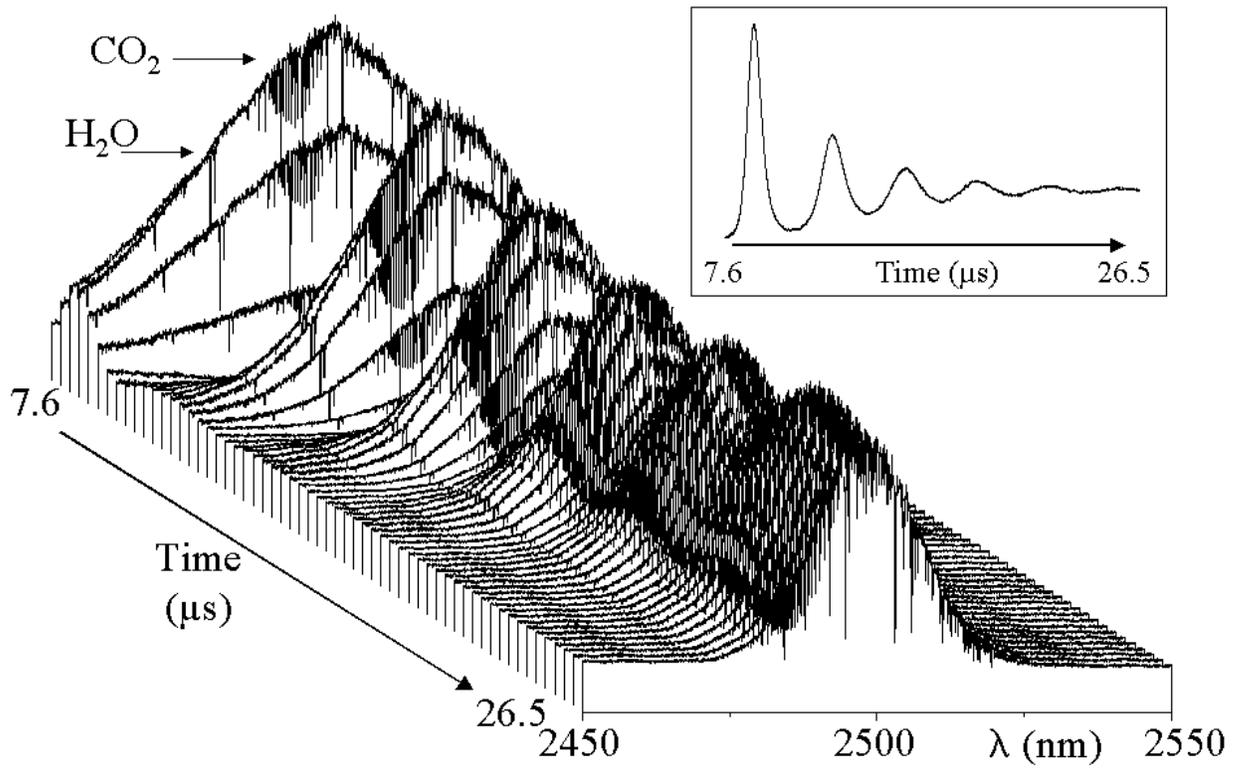

Figure 2

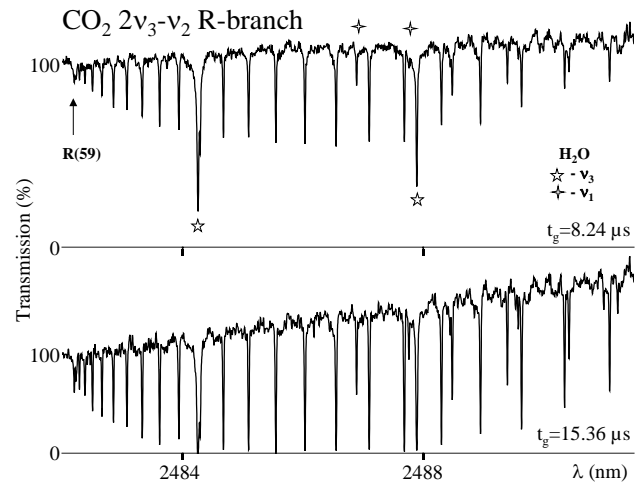

Figure 3